\documentclass[11pt]{article}

\usepackage{graphicx}
\usepackage{endfloat}
\usepackage{amssymb}
\usepackage{amsmath, mathtools, bbm, bm}
\mathtoolsset{showonlyrefs}
\usepackage{subcaption}
\usepackage{natbib}
\usepackage{JASA_manu}
\bibpunct{(}{)}{;}{a}{}{,}

\graphicspath{ {./Figures/}}

\begin{document}

\title{Bayesian Change Point Analysis of Linear Models on Graphs}
\author{Xiaofei Wang\textsuperscript{1} and John W. Emerson\textsuperscript{2}\\ \\
\textsuperscript{1}Department of Mathematics and Statistics\\
Amherst College, Amherst, MA 01002\\ \\
\textsuperscript{2}Department of Statistics \\ 
Yale University, New Haven, CT 06511  \\ \\
email: \texttt{swang@amherst.edu} }

\maketitle

\newpage

\mbox{}
\vspace*{2in}
\begin{center}
\textbf{Author's Footnote:}
\end{center}
Xiaofei Wang is Lecturer,
Department of Mathematics and Statistics, Amherst College, Amherst, MA 01002 (E-mail: swang@amherst.edu). John W. Emerson is Director of Graduate Studies and Associate Professor Adjunct, Department of Statistics, Yale University, New Haven, CT 06511 (E-mail: john.emerson@yale.edu). This work was partially supported by the Yale University Biomedical High Performance Computing Center, funded by NIH grants RR19895 and RR029676-01.

\newpage
\begin{center}
\textbf{Abstract}
 
\end{center}
Consider observations $y_1,\dots,y_n$ on nodes of a connected graph, where the $y_i$ independently come from $N(\theta_i, \sigma^2)$ distributions and an unknown partition divides the $n$ observations into blocks. One well-studied class of change point problems assumes the means $\theta_i$ are equal for all nodes within contiguous blocks of a simple graph of sequential observations; both frequentist and Bayesian approaches have been used to estimate the $\theta_i$ and the change points of the underlying partition. This paper examines a broad class of change point problems on general connected graphs in which a regression model is assumed to apply within each block of the partition of the graph. This general class also supports multivariate change point problems. We use Bayesian methods to estimate change points or block boundaries of the underlying partition. This paper presents the methodology for the general class of change point problems and develops new algorithms for implementation via Markov Chain Monte Carlo. The paper concludes with simulations and real data examples to demonstrate application of the methodology on a wide range of problems. 

\vspace*{.3in}


\newpage

\section{Introduction}
The classical change point problem considers sequential observations $y_1,\dots,y_n$, assumed to come from $N(\theta_i, \sigma^2)$ distributions. It assumes that an unknown partition divides the $n$ observations into blocks. The goal is to recover the $\theta_i$ and, perhaps, to infer the location of change points given the data. There are many methods that handle this classical change point problem and its close relatives; see, for example, \cite{chen2012parametric}. The original Bayesian approach to this problem was developed in \cite{barry93}, which models the unobserved partition using a product partition distribution \citep{hartigan1990partition}. 

This article proposes new Bayesian methodology, also using product partition distributions, that tackles a more general change point problem. We consider observations $\left\{(\bm{x_i}, y_i)\right\}_{i=1}^n$ residing at $n$ nodes of a connected graph and assume an unknown underlying partition divides the nodes into blocks sharing the same distribution parameters. Two nodes joined by an edge are more likely to be in the same block than are non-neighboring nodes, although nodes of a block need not be contiguous. We consider a linear model regressing $y$ on $\bm{x}$ within each block. This general methodology, fully specified in Section \ref{sec:extensions}, is sufficiently flexible to support many different forms of change point analysis.

Special cases encompassed by this generalized change point problem have been studied previously. In most instances, as in the classical change point problem studied by, for example, \cite{erdman2007bcp,Erdman01102008}, observations are assumed to be sequential; this implies a path graph as the underlying structure. The observations can also be multivariate, sharing a common change point structure across all dimensions \citep{perreault2000retrospective,zamba2006multivariate,zamba2009multivariate}. Some approaches additionally allow for changes in variance \citep{james2013ecp,killick2012optimal,killick2011changepoint}. Popular areas of application include the environmental sciences \citep{perreault2000retrospective,chen2012parametric} and finance \citep{holbert1982bayesian,lavielle2006detection,james2013ecp}. 
Another variation uses serial observations to fit linear models within blocks \citep{zeileis2001strucchange,JAE:JAE659,muggeo2003estimating,muggeo2008segmented,fearnhead2005exact,seidou2007bayesian,loschi2010multiple}. Applications are found in econometrics \citep{loschi2010multiple}, environmental sciences \citep{qian2006estimating}, and biomedical sciences \citep{muggeo2003estimating,muggeo2010efficient}. And motivated by the application of image restoration, \cite{barryimage} presented a Bayesian methodology for conducting classical change point analysis on a grid graph. We will show that many of these change point problems are special cases of our general form of the problem. 

We present the theoretical construction and implementation of our methodology in Section \ref{sec:extensions}; selected relevant derivations appear in Appendix A. We show that our methodology encompasses (but is not limited to) several important constant-variance change point problems, including multivariate change point analysis for sequential data, univariate change point analysis for data on a grid graph, and linear regression change point analysis for data on a general graph. Section \ref{sec:simulations} provides simulation results comparing available methodologies on grid graph problems. Most significantly, the new BCP-Graph-0 algorithm demonstrates robustness with respect to the often difficult choice of an important tuning parameter. Section \ref{sec:simulations} also applies the methodology to real data problems. Section \ref{sec:discussion} offers a brief discussion. Our methodology is newly implemented in a major revision of \verb+R+ package \verb+bcp+ (version 4.0.0).

\section{Methodology} 
\label{sec:extensions}
\subsection{Model} 
\label{sub:model}
We consider a connected graph with $n$ nodes. An observation $(\bm{x_i}, y_i)$ is recorded at the $i$-th node, where $\bm{x_{i}}=(x_{i1},\dots,x_{ik})$ is a vector of values for $k$ predicting variables.
Given a partition $\rho$ dividing all nodes into $b$ blocks, each block $S$ is associated with a vector $y_S$ and a matrix $X_S$; $y_S$ contains all observed $y_i$ for nodes $i$ in block $S$ and $X_S$ contains the corresponding $\bm{x_i}$ as rows. 
For a block containing $n_S$ observations, denote $\widetilde{X}_S$ to be the $n_S \times (k+1)$ matrix with a column of ones prepended to $(I_{n_S}-1/n_S J_{n_S})X_S$, a transformed design matrix with each predicting variable centered about its blockwise mean. $I_p$ is the $p\times p$ identity matrix and $J_p$ is the $p\times p$ matrix of ones. 
We assume $y_S\sim N_{n_S}(\widetilde{X}_S\bm{\gamma_S}, \sigma^2 I_{n_S})$ with $\bm{\gamma_S}=(\alpha_S, \beta_{S1},\dots,\beta_{Sk})^\top$ and unknown variance $\sigma^2$. The prior for intercept $\alpha_S$ is
    \begin{align}
      \alpha_S &\sim N\left(\alpha_0, \frac{\sigma_0^2}{n_S}\right), \label{eq:intercept-prior}
    \end{align}
where $\alpha_0 \sim U(-\infty, \infty)$. Without any constraints on the size of a block in a partition, it is possible for a block to have too few observations to fit the model coefficients.  Rather than assigning zero mass in the prior for all such partitions $\rho$ having at least one small block, our model fits only an intercept in small blocks (having $2k$ or fewer observations). For other blocks, the prior is a mixture of the intercept-only model and the full regression model. A parameter $\tau_S$ accomplishes this via the following prior:
        \begin{align}
           P(\tau_S=0) &=\frac{d}{n_S+d}\mathbbm{1}\{n_S\geq 2k\} + \mathbbm{1}\{n_S<2k\}\;\;\;\;\;\mbox{ and}\\
           P(\tau_S=1) &= \frac{n_S}{n_S+d} \mathbbm{1}\{n_S\geq 2k\}.
        \end{align}
  If $\tau_S=0$, then our model assumes only an intercept within block $S$ and, consequently, $\beta_{Sj}=0$ for all $j$; if $\tau_S=1$, then we use the full linear model for block $S$ and adopt the following prior on the regression parameters: $\beta_{Sj} \sim N\left(0,\sigma_j^2/V_{Sjj}\right)$, 
  where $V_{Sjj}$ is the $(j,j)$-th element of $V_S= \widetilde{X}_S^\top\widetilde{X}_S$. We use a different $\sigma_j^2$ for each predictor to account for changes in variance across different predictors. Instead of imposing priors on each $\sigma_j^2$ directly, we consider priors on the error-to-signal ratios 
    $w_j=\frac{\sigma^2}{\sigma^2+\sigma_j^2}\sim U(0, w_j')$ for $j=0,\dots,k$. The $w_j'$ are hyperparameters requiring user input; $w_j'=0.2$ for all $j$ works well in most settings. 
    In addition, we use an improper prior on the error variance:
        $\pi(\sigma^2) \propto 1/\sigma^2$, where $\sigma^2\in(0,\infty)$.
    Finally, we use the following prior on the partition $\rho$: 
    \begin{align}
        \pi(\rho)&\propto \alpha^{l(\rho)},\label{eq:bcp-graph-rho-prior}    
    \end{align}
    where $\alpha < 1$ is a pre-specified parameter and $l(\rho)$ is the total boundary length of the partition, calculated as $\sum_{j=1}^b l(S_j)$, with $l(S_j)$ as the number of nodes that are neighbors of $S_j$. A neighbor of block $S_j$ is not in $S_j$ but shares at least one edge with a node in $S_j$. This prior is similar to the prior given in \cite{barryimage} for classical change point analysis on a grid graph and shares the same ``short boundary'' property that encourages the number of adjacent nodes belonging to different blocks to be small. Intuitively, $\alpha$ represents the multiplicative penalty in the likelihood for each additional boundary edge. $\alpha<1$ implies a preference for shorter boundaries over longer ones; the smaller the $\alpha$, the heavier the penalty for each boundary edge.

  \subsection{Conditional Distributions and Expectations} 
  \label{sub:posterior_distributions}
   We begin with some notation before presenting the relevant formulas used in implementing our methodology. Given a partition, let $W=\sum_S \sum_{i\in S} (y_{i}-\bar y_S)^2$ and $B=\sum_S n_S (\bar y_S - \bar y)^2$ be the within-block and between-block sums of squares, respectively, where $\bar y_S$ is the mean of $y_i$ within block $S$ and $\bar y$ is the overall mean of all $y_i$. Define 
      \begin{align}
           Z_S=\left(\begin{array}{ccccc}
        \frac{1-w_0}{n_S w_0} & 0 & 0 &\dots &0\\
        0 & \frac{1-w_1}{V_{S11}w_1} & 0 &\dots &0\\
        \vdots & \vdots & \vdots & \vdots &\vdots\\
        0 & 0 & \dots & 0 & \frac{1-w_k}{V_{Skk}w_k}
        \end{array}\right).
    \end{align}
    For a matrix $Q$, denote $Q_{(-1,-1)}$ as the submatrix formed by removing the first row and first column from $Q$, and denote $Q_{-1}$ as the submatrix formed by removing the first row from $Q$. Let $\bm{\hat\beta_S}$ be the posterior mean of the slope coefficient(s) in block $S$ given $\bm{w}$ and $\tau_S=1$, where $\bm{\hat\beta_S}=\left[(\widetilde{X}_S^\top\widetilde{X}_S+Z_S^{-1})^{-1}\widetilde{X}_S^\top y_S\right]_{-1}$. Finally, let $\widetilde{W}  = W - \sum_{S:\tau_S=1}\bm{\hat\beta_S^\top}(\widetilde{X}_S^\top\widetilde{X}_S+Z_S^{-1})_{(-1,-1)}\bm{\hat\beta_S}$. 

    This model leads to a number of conditional distributions (derived in the Appendix) used in the implementation. For $B>0$,
        \begin{align}
        f(\rho, \tau|y, \bm{x}, \bm{w}) 
          \propto&  f(\tau|\rho)f(\rho) f(y|\bm{x}, \rho, \bm{w}, \tau) \\
          \propto&\prod_{S}\left[\left(\frac{d}{n_S+d}\mathbbm{1}\{n_S\geq 2k\} + \mathbbm{1}\{n_S<2k\}\right)\mathbbm{1}\{\tau_S=0\}\right.\\
          &\quad\quad\left.+\left(\frac{n_S}{n_S+d}\mathbbm{1}\{n_S\geq 2k\}\right)\mathbbm{1}\{\tau_S=1\}\right]\\
          &\times\alpha^{l(\rho)} \\          
          &\times
          \frac{w_0'^{(b-1)/2}\prod_{S:\tau_S=1}\left|(\widetilde{X}_S^\top\widetilde{X}_SZ_S +\ I)_{(-1,-1)}\right|^{-1/2}}{B^{(b+1)/2} \widetilde{W}^{(n-b-2)/2}}\\
          &\times Beta\left(\frac{Bw_0'/\widetilde{W}}{1+Bw_0'/\widetilde{W}}; \frac{b+1}{2}, \frac{n-b-2}{2}\right)\;\;\;\;\;\mbox{ and}\label{eq:bcpr-rhotau-post} \\
      f(\bm{w}|y, \bm{x}, \rho, \tau) 
        \propto& f(y|\bm{x}, \rho,\bm{w}, \tau) f(\bm{w}) \\
       \propto& \frac{w_0'^{(b-1)/2}\prod_{S:\tau_S=1}\left|(\widetilde{X}_S^\top\widetilde{X}_SZ_S + I)_{(-1,-1)}\right|^{-1/2}}{B^{(b+1)/2} \widetilde{W}^{(n-b-2)/2}}\\
          &\times Beta\left(\frac{Bw_0'/\widetilde{W}}{1+Bw_0'/\widetilde{W}}; \frac{b+1}{2}, \frac{n-b-2}{2}\right).\label{eq:bcpr-w2-post}
    \end{align}
    If instead $B=0$ (which certainly occurs when $b=1$ but might even occur with real data in non-trivial cases), then
    \begin{align}
    f(\rho, \tau|y, \bm{x}, \bm{w}) 
          \propto&\prod_S\left[\left(\frac{d}{n_S+d}\mathbbm{1}\{n_S\geq 2k\} + \mathbbm{1}\{n_S<2k\}\right)\mathbbm{1}\{\tau_S=0\}\right.\\
          &\quad\quad\left.+\left(\frac{n_S}{n_S+d}\mathbbm{1}\{n_S\geq 2k\}\right)\mathbbm{1}\{\tau_S=1\}\right]\\
          &\times\alpha^{l(\rho)}  \\          
          &\times
          \frac{w_0'\left|(\widetilde{X}_S^\top\widetilde{X}_SZ_S + I)_{(-1,-1)}\right|^{-1/2}}{ \widetilde{W}^{(n-1)/2}}\;\;\;\;\;\;\mbox{ and}\label{eq:bcpr-rhotau-post1}\\
      f(\bm{w}|y, \bm{x}, \rho, \tau) 
       \propto& \frac{w_0'\left|(\widetilde{X}_S^\top\widetilde{X}_SZ_S + I)_{(-1,-1)}\right|^{-1/2}}{\widetilde{W}^{(n-1)/2}}.\label{eq:bcpr-w2-post1}
    \end{align}   
    Define $w_0^*=E(w_0|y,\bm{x},\rho,\tau)$. The conditional expectations, given below, are used for estimating the posterior means:
 \begin{align}
E(\alpha_S|y,\bm{x}, \rho, \bm{w}, \tau) &= (1-w_0^*) \bar y_S + w_0^* \bar y,\label{eq:bcpr_alpha_post}\\
    E(\bm{\beta_S}|y, \bm{x}, \rho, \bm{w}, \tau) &= \bm{\hat\beta_S},\;\;\;\;\;\mbox{ and} \label{eq:bcpr_beta_post}\\
    E(\sigma^2|y, \bm{x}, \rho, \bm{w}, \tau) 
    &= \frac{\widetilde W + w_0^* B}{n-3}. \label{eq:bcpr_sig_post}
 \end{align}
  For $B=0$, it is easy to see that $w_0^* = w_0'/2$ and for $B > 0$,
  \begin{align}
  w_0^* &= \frac{\widetilde{W}}{B}\frac{Beta\left(\frac{Bw_0'/\widetilde{W}}{1+Bw_0'/\widetilde{W}}; \frac{b+3}{2}, \frac{n-b-4}{2}\right)}{Beta\left(\frac{Bw_0'/\widetilde{W}}{1+Bw_0'/\widetilde{W}}; \frac{b+1}{2}, \frac{n-b-2}{2}\right)}.
  \end{align}

\subsection{Implementation via Markov Chain Monte Carlo}
  Due to the complexity of the posterior distributions, we use Markov Chain Monte Carlo (MCMC) to sample the partition $\rho$ and $\tau$. Specifically, we use a combination of pixel passes adapted from \cite{barryimage} and two new passes. We describe each of these below.

  The pixel passes make use of \eqref{eq:bcpr-rhotau-post} and \eqref{eq:bcpr-rhotau-post1} for sampling. An All (Full) Pixel Pass (FPP) iterates over each node in turn, allowing the current node to join any one of the other $b-1$ blocks, stay in its current block, or form a new block while simultaneously updating $\tau_S$ for the affected block(s). FPPs are computationally intensive, requiring approximately $O(nb)$ likelihood calculations at each pass (where $b$ may change during the pass). 

  Active Pixel Passes (APPs) take advantage of the fact that nodes interior to a block are unlikely to switch memberships because of the heavy likelihood penalty that would be incurred for new boundary edges. By construction, APPs skip nodes that are in the interior of a block and only consider membership updates for nodes that are \textit{active}, residing at the boundary of a block. If node $i$ is active, then one of two things happen: 1) if node $i$ agrees with at least one other neighbor, then the node may join a neighboring block and 2) if, instead, node $i$ is an island (disagreeing with all of its neighbors), then it is only allowed to join blocks that preserve its island status. Again, as we determine the new block membership for node $i$, we simultaneously consider updating $\tau$ for the affected block(s). The preservation of island nodes is a key requirement in the conditioned transition argument that justifies APPs \citep{barryimage} while simultaneously being at odds with the short boundary philosophy. Our simulations in Section \ref{sub:change_points_on_a_graph_simulations} suggest that the treatment of island nodes in the APP causes the number of blocks $b$ to be extremely sensitive to $\alpha$; when $\alpha$ is slightly larger than the acceptable range, the sampled partitions have too many island nodes, increasing the runtime and resulting in noisy posterior means. 

  We propose a pseudo-APP that modifies the APP only in its treatment of island nodes. A pseudo-APP treats all active nodes, island or otherwise, the same way, allowing the node to join any of its neighboring blocks or stay in its current block. Simulations suggest that using the pseudo-APP in place of the original APP increases robustness to $\alpha$, widening the acceptable range of $\alpha$ values significantly (Section \ref{sub:change_points_on_a_graph_simulations}). Unfortunately, the modified pseudo-APP breaks the conditioned transition argument and is no longer a legal MCMC move. Although the resulting sample is not guaranteed to come from the posterior distribution $f(\rho, \tau|y)$, our simulations show that for short-boundary scenes, pseudo-APPs produce estimates that are more in line with the ground truth than are estimates produced by original APPs. To make a fair comparison, we will also illustrate some scenarios where the pseudo-APPs fare worse. Our software implementation allows the user to specify the proportion of APPs that are pseudo-APPs. 

  In addition to the pixel passes, we also use a block merge pass and a $w$-pass. The block merge pass iterates over each block $S_j$ for $j=1,\dots,b$. For block $S_j$, considering merging $S_j$ with another block $S$ in the current partition satisfying $\tau_S = \tau_{S_j}$. The $w$-pass allows for iteratively updating $w_l$ for $l=1,\dots,k$ using \eqref{eq:bcpr-w2-post} and \eqref{eq:bcpr-w2-post1}.

  The algorithm begins with 100 FPPs that are discarded as part of the burn-in. We then proceed with $M$ additional steps, each consisting of 1 FPP, 20 APPs, 1 block merge pass, and 1 $w$-pass; some of these steps may be additionally discarded as part of the burn-in. 
  After discarding a number of iterations for the burn-in, each pass through the data ends in a calculation of the conditional expectations given by \eqref{eq:bcpr_alpha_post}, \eqref{eq:bcpr_beta_post}, and \eqref{eq:bcpr_sig_post}. Given a partition $\rho$ at the $t$-th MCMC step, we calculate a posterior mean $\hat y_{it}$ for observation $i$ in block $S$. 
  Finally, we obtain our estimates for the posterior means $\{\hat y_i\}$ and its associated variances by aggregating over the conditional expectations calculated in each pass.

  Note that when all $x_{ij}$ are equal within a block $S$ for a predictor $j$, a singularity in $\widetilde{X}_S^\top\widetilde{X}_S$ renders certain calculations in the algorithm impossible. This can happen, for example, if variable $j$ is discrete-valued. 
    In such situations, our algorithm temporarily adds a small amount of noise to the data for the calculation in which the singularity is encountered. The noise is independently regenerated for subsequent calculations when necessary.

\subsection{Special Cases} 
\label{sub:special_cases}
  Classical change point analysis uses sequential observations, implying a path graph with a partition of consecutive blocks. In this setting, blocks differ in the mean parameter of the underlying normal distributions. Only fitting a mean within each block is equivalent to fitting trivial intercept-only linear models within each block. Therefore, classical change point is a special case of the change point problem on a graph with linear regression.

  When the underlying graph structure is a path graph, we recommend borrowing elements from \cite{barry93}, replacing the partition prior \eqref{eq:bcp-graph-rho-prior} with $\pi(\rho) = \int_0^{p_0} p^{b-1} (1-p)^{n-b}dp$ for some prespecified $p_0$. This prior reflects the assumption that each node is equally likely to be a boundary node with equal probability $p$. The MCMC algorithm simplifies greatly with a path graph; rather than making MCMC moves that alter the block membership node by node, we instead consider breaking or merging blocks at each possible change point location.  

  In multivariate change point analysis, we fit multivariate normal distributions, thereby estimating a vector of means within each block. We can accomplish this in the general change point framework by realizing that fitting a vector of means is equivalent to performing a one-way analysis of variance. To conduct $k$-dimensional change point analysis for $n$ observations, we view our observed data at node $i$ as $\bm{y_i}=(y_{i1},\dots,y_{ik})^\top$ and $\bm{x_i}=I_k$. Since all of the coefficients $\alpha_1,\dots,\alpha_k$ in this model are  means, we directly use the design matrix $X_S$ without centering and adopt the intercept prior \eqref{eq:intercept-prior} for all $\alpha_j$. Moreover, because there are no coefficients to estimate, we effectively set $\tau_S=0$ for all blocks $S$ in the partition, and there is no need to sample $\tau$. 
  In the same vein, our method can be generalized to include multiple observations at each node, even in the regression setting.

\section{Simulations and Applications} 
\label{sec:simulations}
  \subsection{Multivariate versus Univariate Change Point Analysis} 
  \label{sub:multivariate_vs_univariate_change_point}
\begin{figure} 
\begin{center}
  \begin{subfigure}[b]{0.48\textwidth}
    \includegraphics*[width=.95\textwidth]{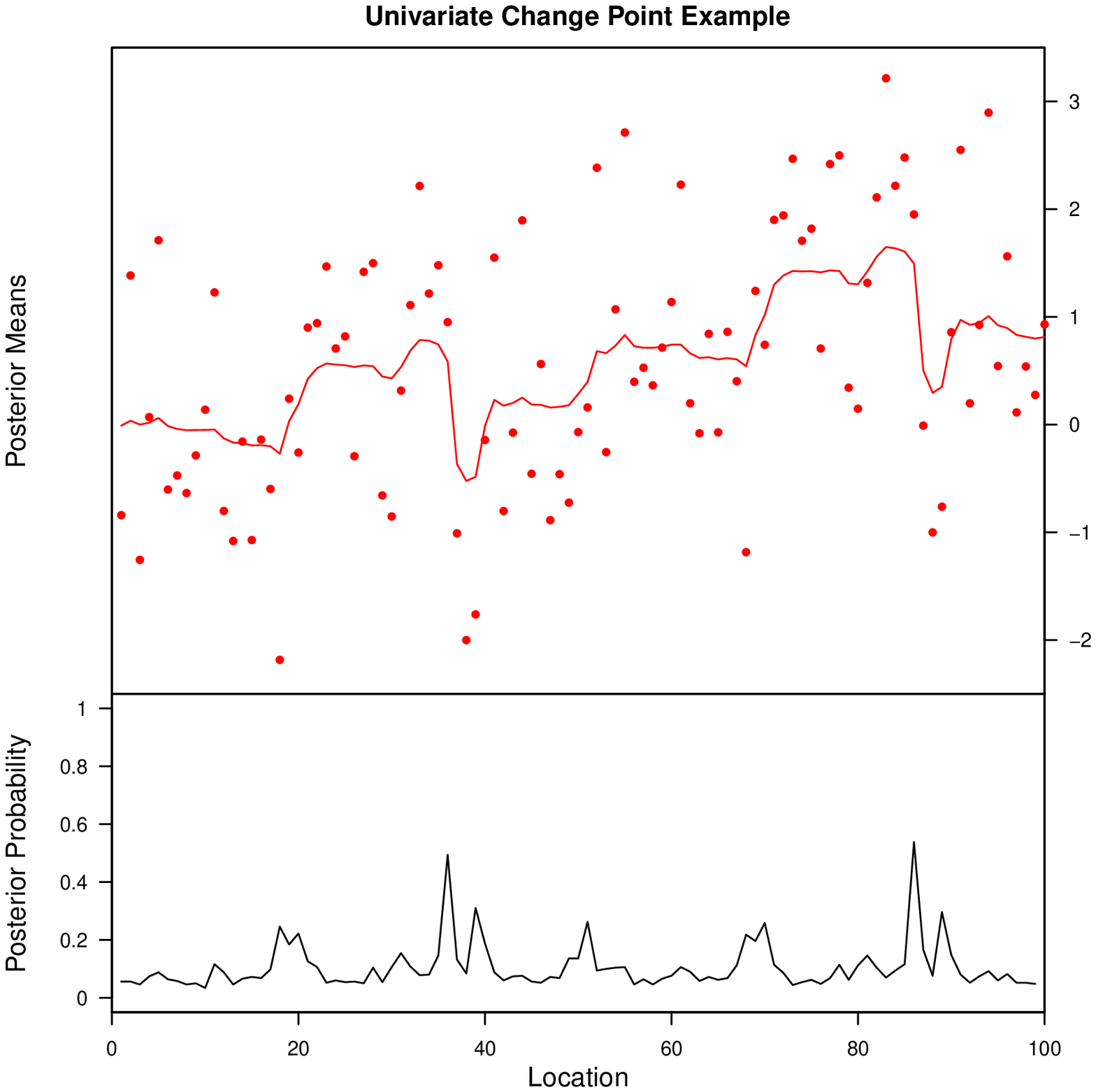}
    \caption{}
  \end{subfigure} ~
  \begin{subfigure}[b]{0.48\textwidth}
    \includegraphics*[width=.95\textwidth]{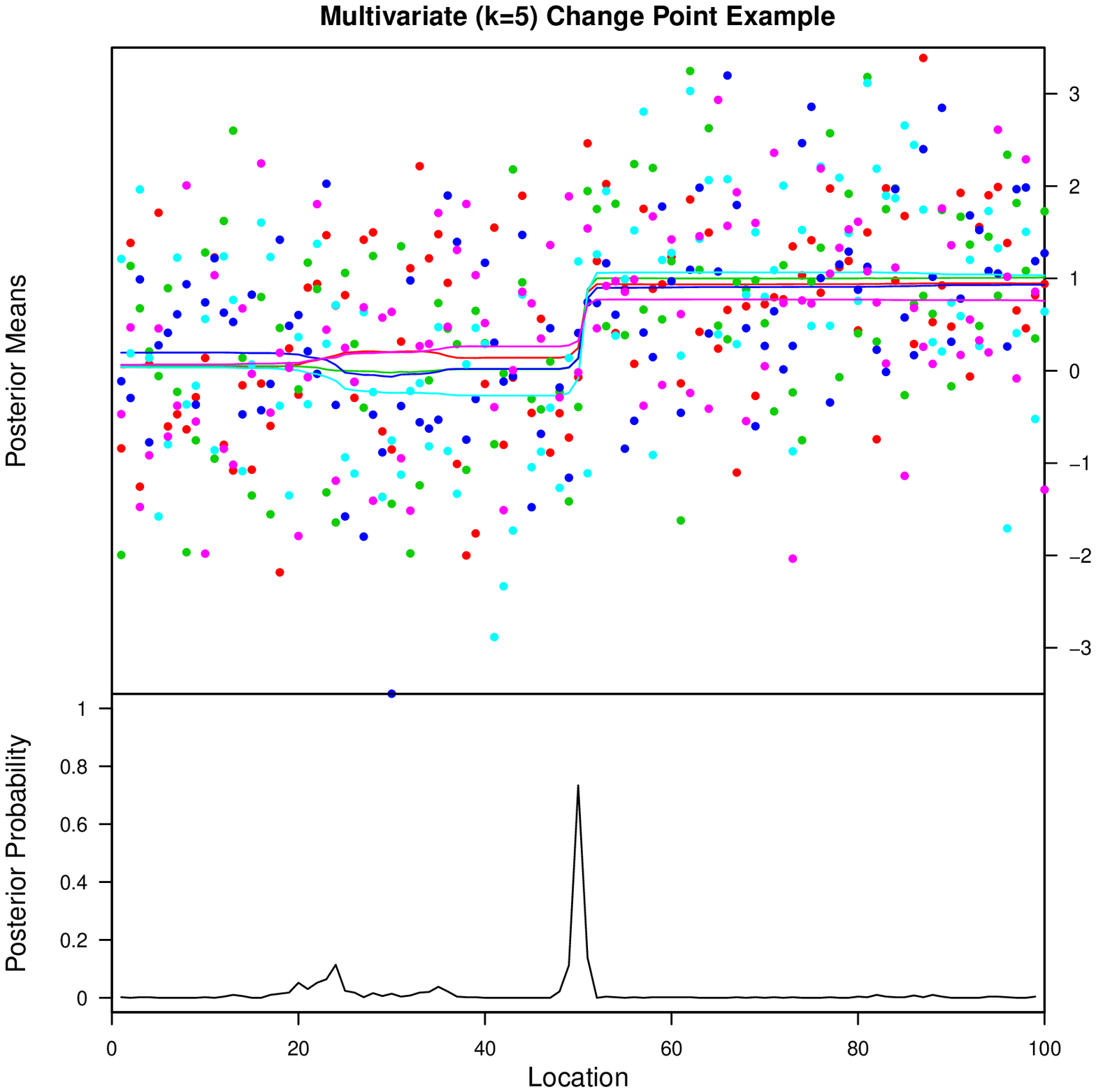}
    \caption{}
  \end{subfigure}
  \caption{Two simulated examples: (a) univariate data $y_i\sim N(\theta_i,1)$, where $\theta_i=0$ for $i\leq 50$ and $\theta_i=1$ for $i>50$; (b) multivariate data $\bm{y_i}\sim N_5(\bm{\theta_i}, I)$, where $\bm{\theta_i}=(0, 0,0,0,0)^\top$ for $i\leq50$ and $\bm{\theta_i}=(1,1,1,1,1)^\top$ for $i > 50$. The blockwise means within each dimension were assumed equal for this particular example, but in general, the blockwise means are unconstrained. The top panels show the simulated data and the posterior means at each location in the series, and the bottom panels show the posterior probability of a change point.}
\label{fig:bcpm-comparison}
\end{center}
\end{figure}

    Multivariate change point analysis is an easier problem than univariate change point analysis; added dimensions yield more information on change point locations because the change points are replicated over each dimension. If the noise level is high, a univariate signal may get lost in the noise far more easily than a multivariate one. Figure \ref{fig:bcpm-comparison} shows the results of our method on a univariate example and a multivariate example sharing the same change point structure. The dataset on the left is one-dimensional and the dataset on the right is five-dimensional. Both datasets in the example have the same underlying change point at $i=50$; the means are 0 and 1 in the first and second blocks across each dimension.
    The algorithm struggles to decisively identify the single change point using the univariate series. In contrast, the multivariate change point results show a  high estimate for the posterior probability of a change point at the true location. 
  
  \subsection{Quebec Streamflow} 
  \label{sub:quebec_streamflow}
\begin{figure}
\begin{center}
\includegraphics*[width=.9\textwidth]{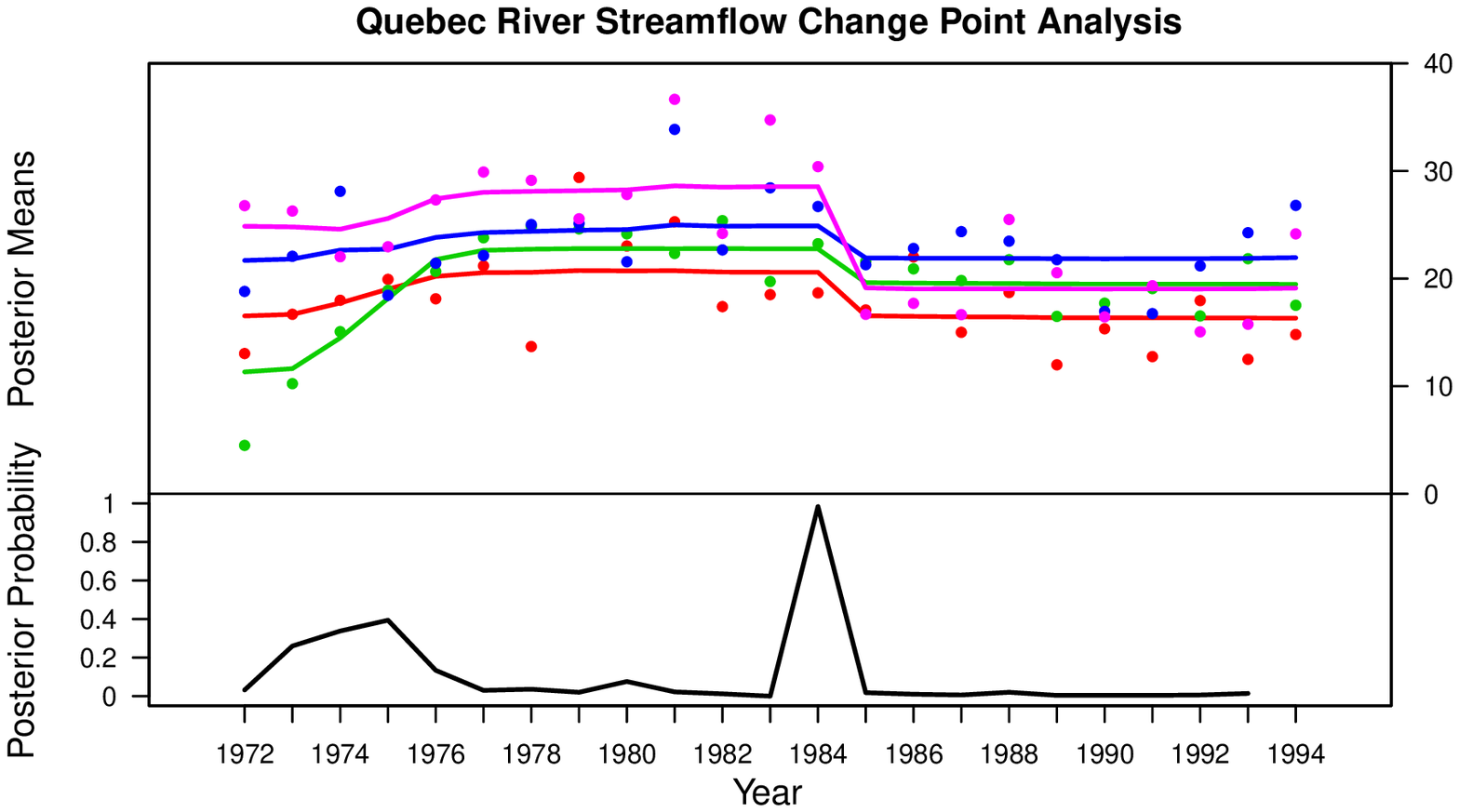}
\caption{Quebec streamflow example. The year 1984 has a high posterior probability of being a change point.}
\label{fig:quebecrivers}
\end{center}
\end{figure}

    \cite{perreault2000retrospective} applied their multivariate single change point detection methodology to annual January to June streamflow (measured in $L/(km^2\cdot s)$) for six rivers in Quebec, Canada. We examine data from four of the six rivers\footnote{http://www.wsc.ec.gc.ca/applications/H2O/index-eng.cfm} in their study; the other two rivers appear in the database but their recorded characteristics differ substantially from the descriptions given in \cite{perreault2000retrospective} and are not included in our example. Using their algorithm, the authors found the year 1984 to be the mode of the marginal posterior distribution of the change point locations, and gave estimates for the means before and after the change point. Our results for the four rivers (Romaine, A la Baleine, Churchill Falls, and Manicouagan) are given in Figure \ref{fig:quebecrivers}. Our method confirms the year 1984 as a likely change point, but additionally discovers possible change points near the beginning of the time period. 

  \subsection{Simulations Involving Change Points on a Grid} 
  \label{sub:change_points_on_a_graph_simulations}
\begin{figure}
  \begin{center}  
      \includegraphics*[width=.9\textwidth]{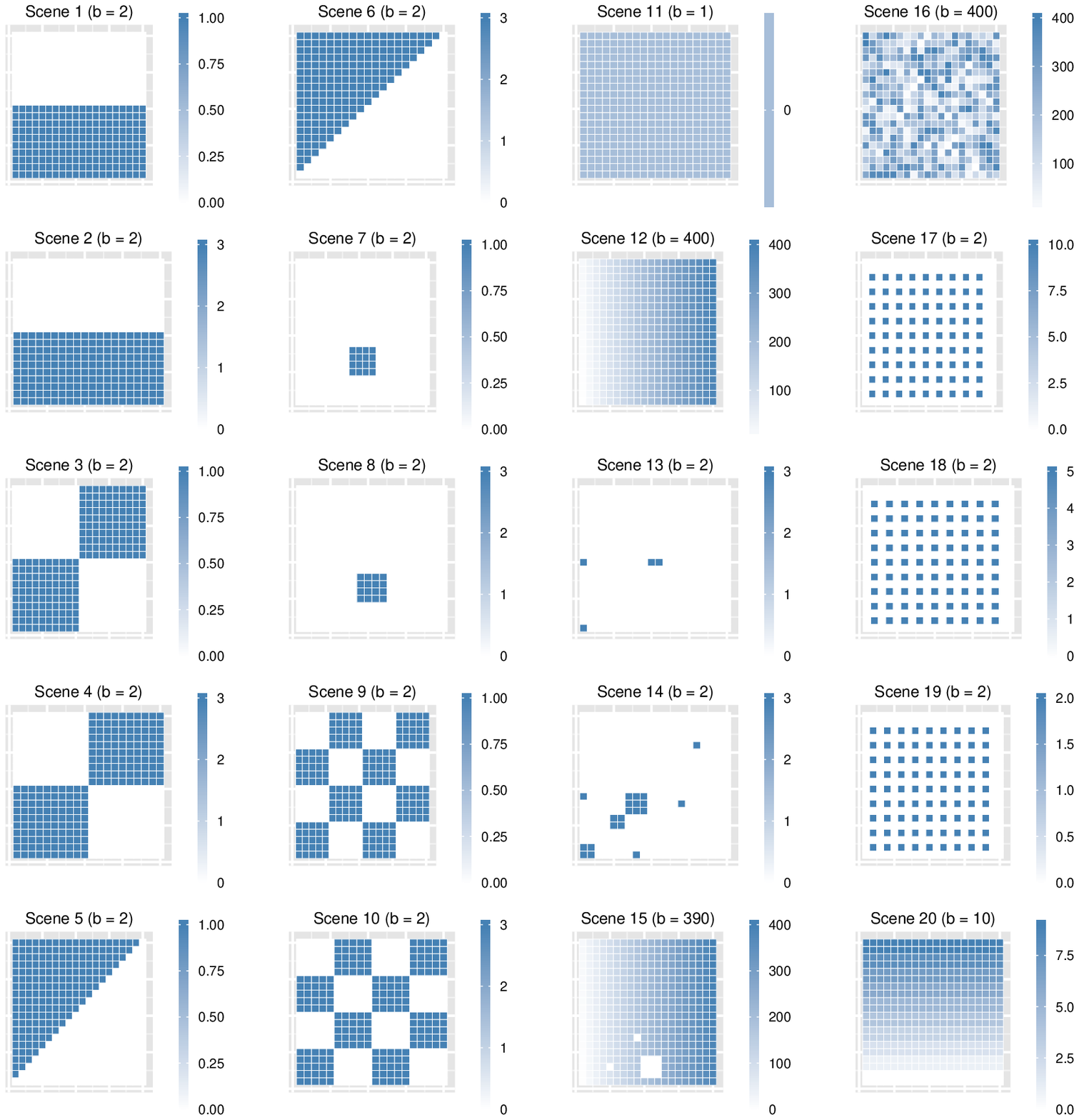}
    \caption{Scene partition boundaries and block means for simulations on 20-by-20 grid graphs. An error variance of $\sigma^2=1$ is used throughout. The true number of blocks $b$ is given in parentheses.}
  \label{fig:gridscenes}
  \end{center}
\end{figure}

    We considered 20 simulated scenes on a 20-by-20 grid graph to compare the method in \cite{barryimage} (BCP-Grid) against our method BCP-Graph; our method assumes up to eight neighbors per node instead of BCP-Grid's four, by additionally including diagonal edges between nodes. These 20 scenes (Figure \ref{fig:gridscenes}) include a range of characteristics representative of real datasets that are appropriate for both methods. For each scene, we simulated 10 datasets and ran each method using a range of values for the parameter $\alpha$. We used $M=2000$ steps with the first 1000 steps discarded as part of the burn-in. 

\begin{figure}
  \begin{center}  
      \includegraphics*[width=.9\textwidth]{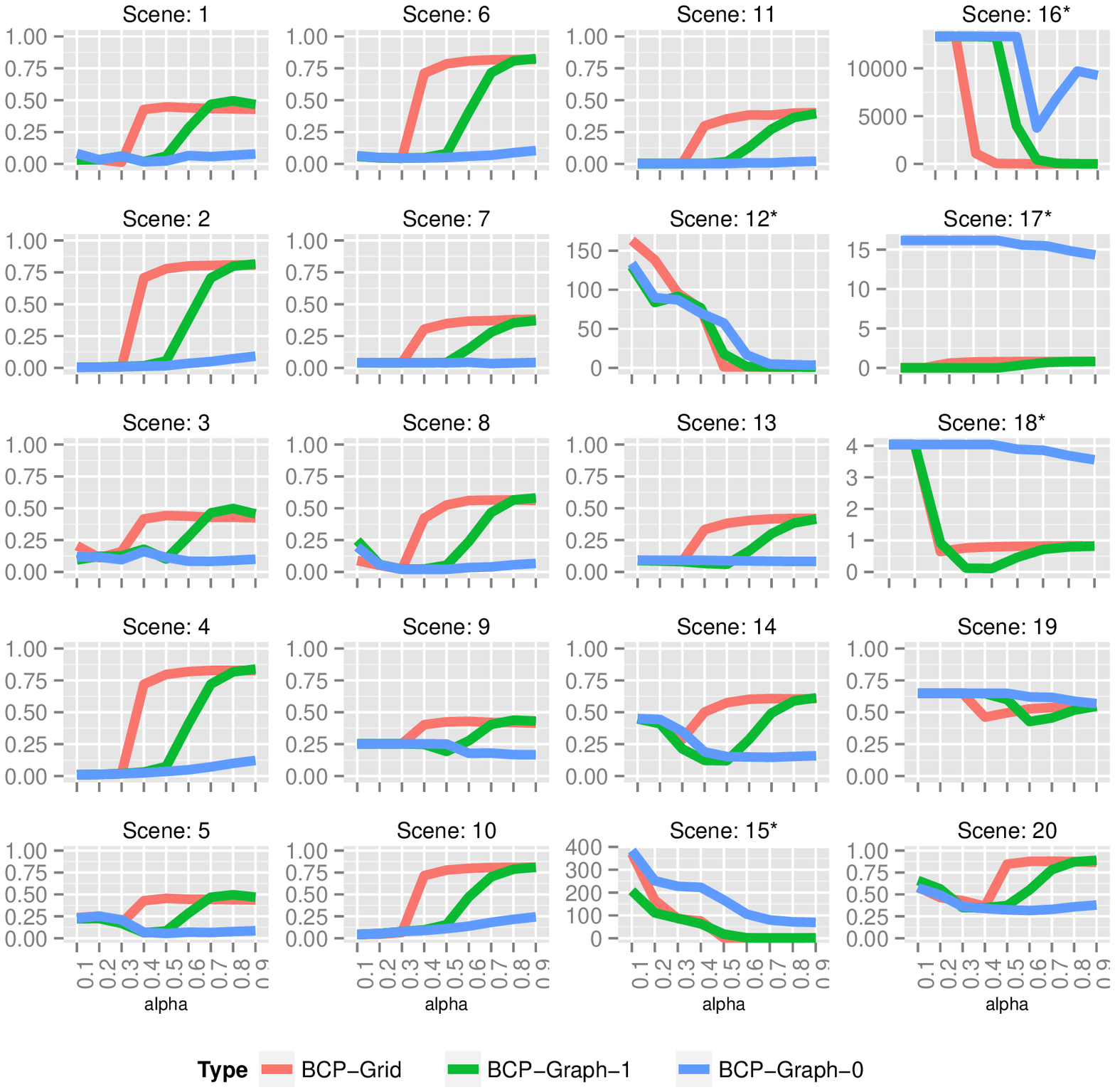}
    \caption{Mean squared error (MSE) for all simulated grid scenes over a large range of values for the parameter $\alpha$. Generally, small values of $\alpha$ produce lower MSE. However, the contrived large-boundary scenes marked with asterisks attain lower MSEs at higher values of $\alpha$.}
  \label{fig:gridscenes-mses}
  \end{center}
\end{figure}

    We plot the mean squared error (MSE) as the main measure of performance in Figure \ref{fig:gridscenes-mses}. BCP-Graph-1 corresponds to our method using Barry and Hartigan's APPs; BCP-Graph-0 corresponds to our method using pseudo-APPs. While comparable levels of MSE are achievable across all three methods for most scenes, BCP-Grid and BCP-Graph-1 both require careful selection of $\alpha$. BCP-Graph-1 has a wider range of acceptable values for $\alpha$ relative to BCP-Grid. Overall, scenes with shorter boundaries (Scenes 1-11) favor lower values of $\alpha$. As $\alpha$ becomes larger, both of these algorithms iterate over partitions with a larger number of blocks, increasing the MSE. Of the three methods, BCP-Graph-0 is the most robust to the choice of $\alpha$ for these shorter boundary scenes, because pseudo-APPs eliminate small isolated blocks in the partition. 
    However, BCP-Graph-0 performs poorly in longer boundary scenes, which have a large number of isolated islands in the underlying true partition (Scenes 15 through 19).  
    
  Plots comparing runtime and mean number of blocks for these simulations can be found in Appendix B. Independently of the choice of $\alpha$, BCP-Graph-0 is far more computationally efficient than either BCP-Grid or BCP-Graph-1 and also encourages partitions having smaller numbers of blocks. 
  
  \subsection{New Haven Real Estate Values} 
  \label{sub:new_haven_real_estate_values}
  We conclude by applying our method to an example modeling housing values.
    Housing values could vary with observed characteristics such as the number of
    bathrooms, number of bedrooms, and living area, or with unobserved characteristics such as neighborhood structure. A single linear model applied to all properties may not be appropriate; different associations between the variables might exist across different neighborhoods. To more accurately
  model housing values over many neighborhoods via linear regression, we would need \textit{a priori} neighborhood boundaries; these may not be available or inaccurately estimated. Our method, however,
  handles the situation gracefully because it considers partitions (neighborhoods) that are not pre-specified and allows slopes and intercepts to vary from one neighborhood to another.

\begin{figure}
  \begin{center}
    \includegraphics*[width=.9\textwidth]{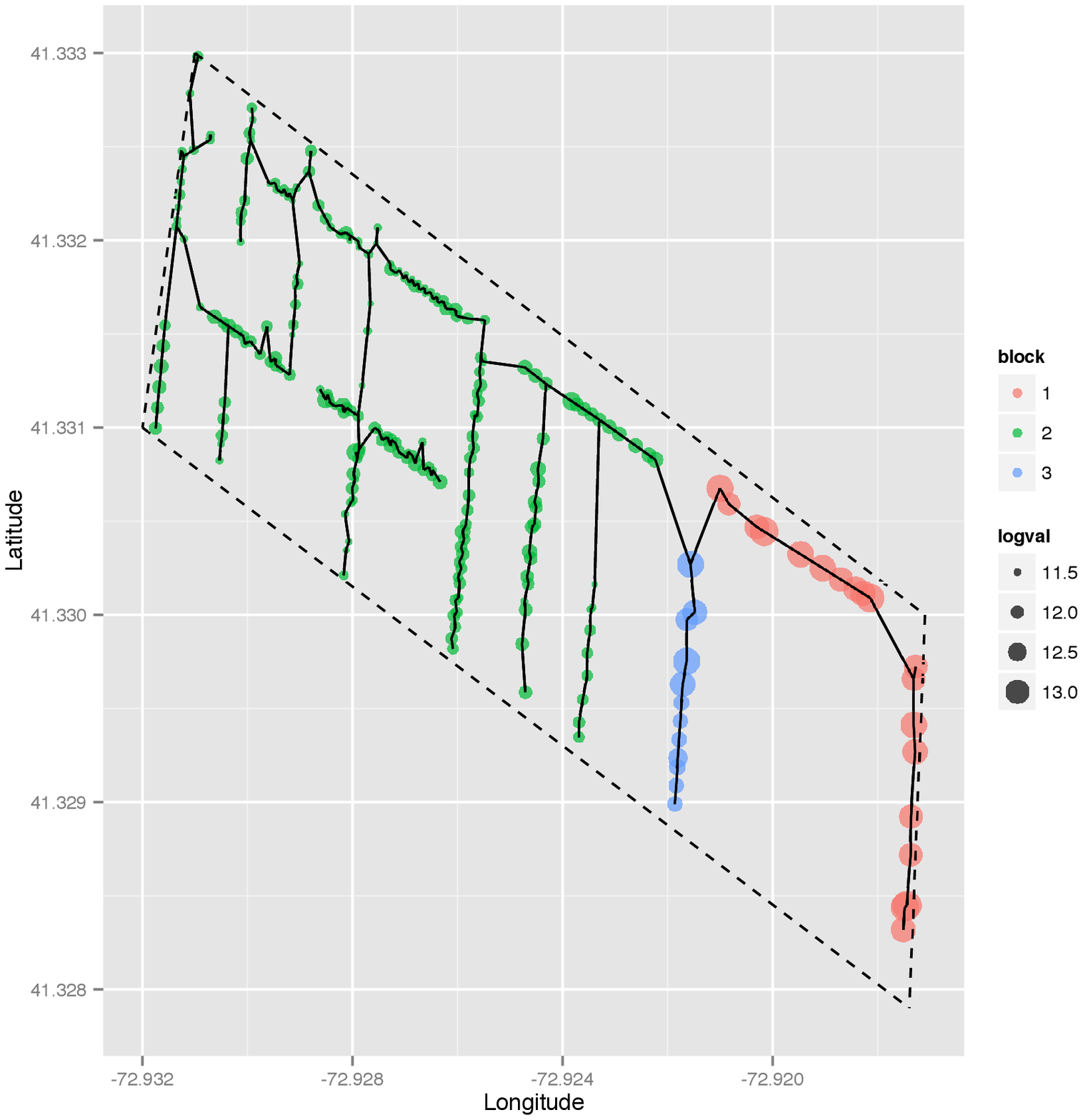}
  \caption{New Haven housing example: minimum spanning tree joining all houses (in circles, with size indicating the magnitude of the true assessed value,  $logval$) with dashed lines demarcating the region of interest, colored by neighborhood membership in the posterior modal partition. Neighborhood 3 houses lie on busy Prospect Street; upscale neighborhood 1 houses lie on quiet Edgehill Road or Huntington Street; neighborhood 2 is a dense neighborhood of generally smaller houses on smaller parcels of land.}
\label{fig:newhavenhousing}
\end{center}
\end{figure}

  Our example uses New Haven, Connecticut residential property data\footnote{http://data.visionappraisal.com/newhavenct/}, and latitudes and longitudes obtained with the Google Maps API. We have limited the data to all 244 houses (excluding condominiums and apartments) within the region outlined by the dashed lines in Figure \ref{fig:newhavenhousing}. We chose this region because it consists of roughly three value-differentiating neighborhoods that are separated with respect to longitude. We model the 2011 log assessed value as a function of the square root of living area, lot size, and number of bedrooms. Although location is not directly used in the linear models, the graph structure and exploration of partitions of the properties into neighborhoods introduces a spatial component to the analysis.

  To use the location of each house in the analysis, we used Euclidean distance on the longitudes and latitudes in generating a minimum spanning tree \citep{prim1957shortest,oksanen2008vegan} over all houses. We then ran our method with $\alpha=0.1$ and $\alpha=0.3$ using pseudo-APPs. 
  The left panel of Figure \ref{fig:newhavenhousing} shows the actual properties with circles, where the size of a circle indicates the magnitude of the actual assessed value. The right panel shows the posterior modal partition (for both $\alpha=0.1$ and $\alpha=0.3$) consisting of three neighborhoods. Upon visual inspection, the large neighborhood 2 off to the west consists of houses that are tightly clustered together and are generally lower-valued than properties in the other neighborhoods. Neighborhood 1 to the east consists of more expensive houses that are more spread apart (having larger lots). Neighborhood 3 separates the other two neighborhoods and seem to be a mix of cheaper and more expensive houses. Both $\alpha=0.1$ and $\alpha=0.3$ produced the same modal partition, but $\alpha=0.1$ resulted in an average partition size of 3.01 neighborhoods and $\alpha=0.3$ resulted in an average partition size of 4.46 neighborhoods. 

\begin{figure}
\begin{center}
  \begin{subfigure}[b]{0.3\textwidth}
    \includegraphics*[width=.95\textwidth]{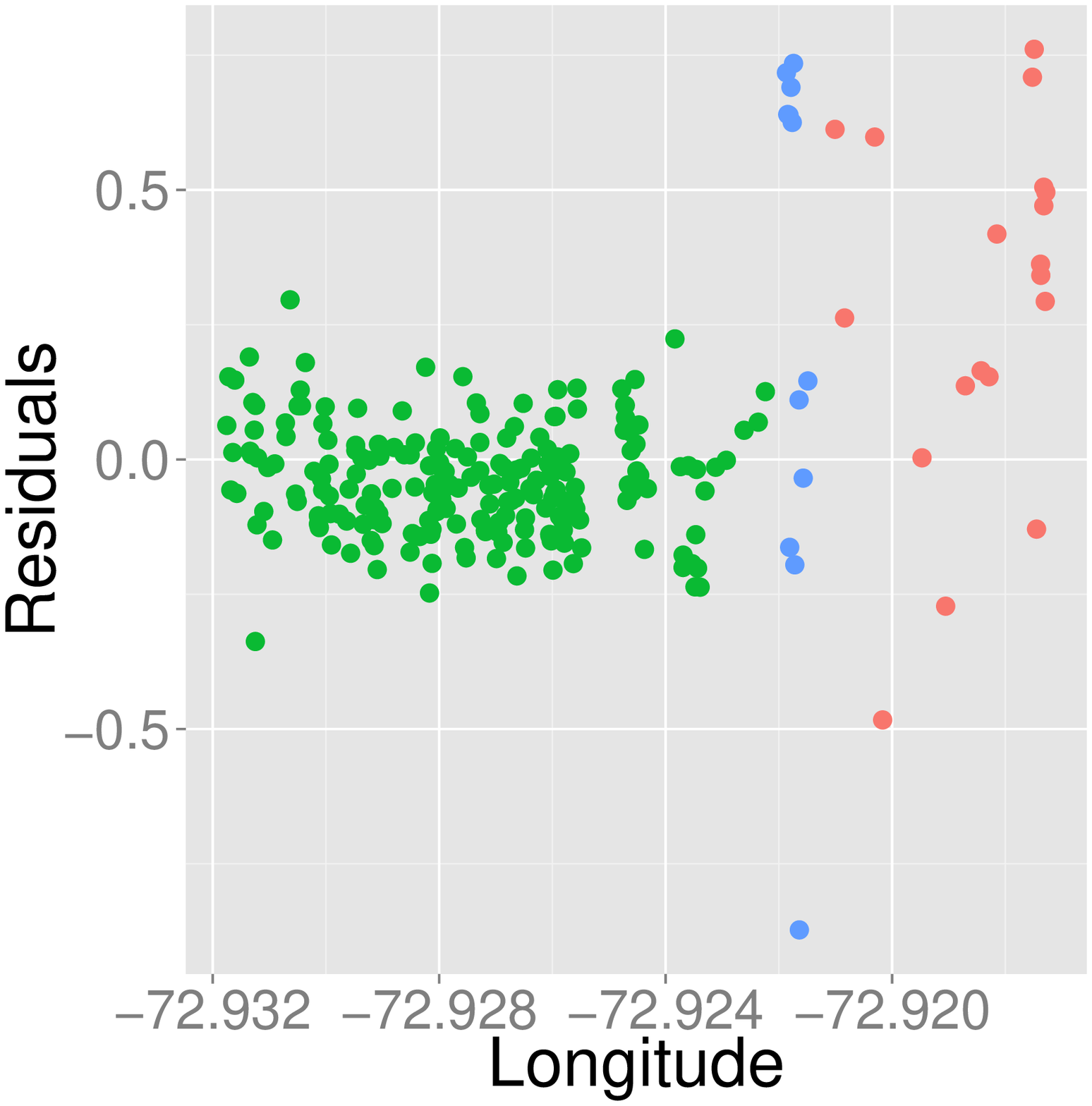}
    \caption{}
  \end{subfigure} ~
  \begin{subfigure}[b]{0.3\textwidth}
    \includegraphics*[width=.95\textwidth]{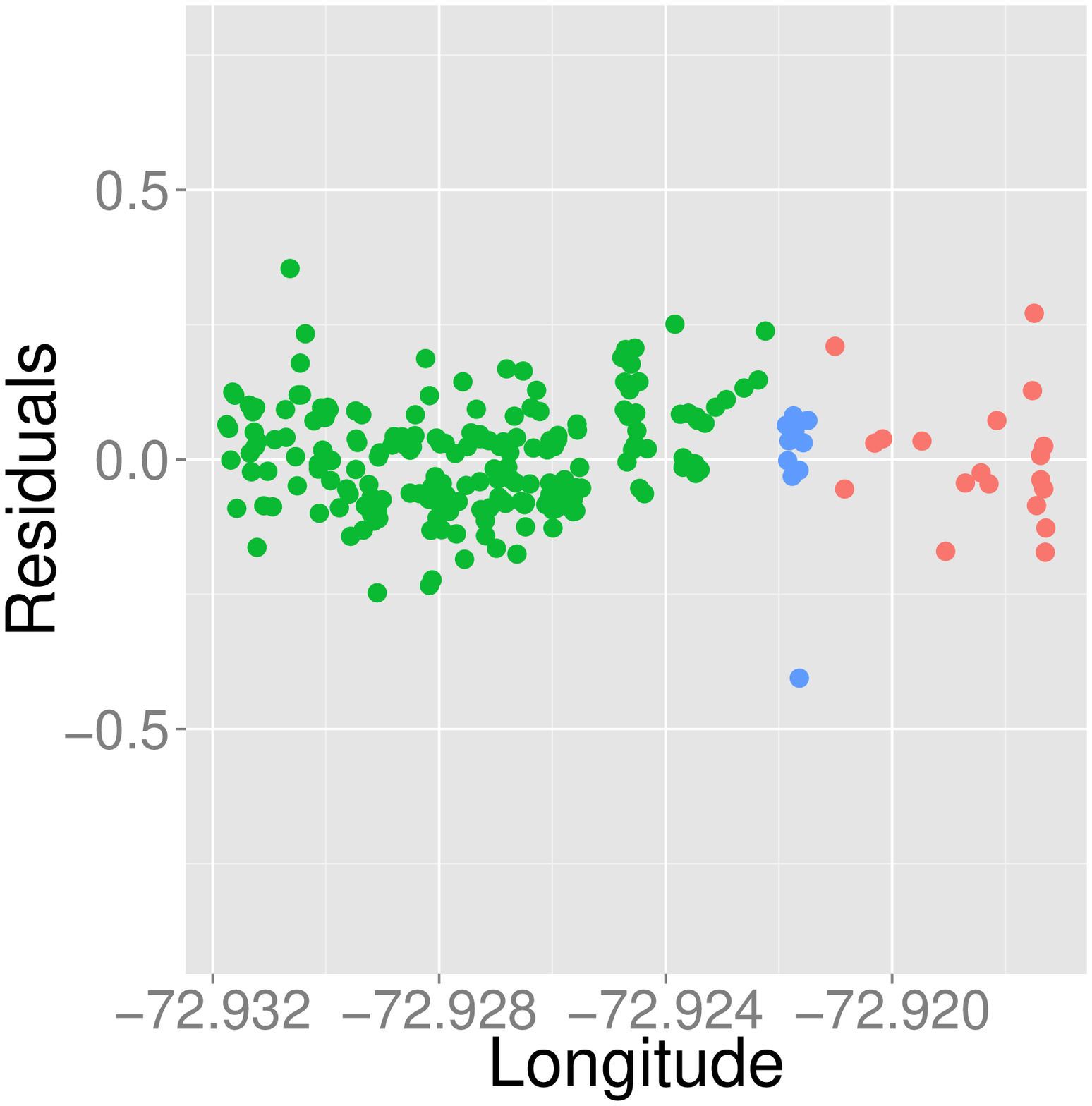}
    \caption{}
  \end{subfigure}
  \begin{subfigure}[b]{0.3\textwidth}
    \includegraphics*[width=.95\textwidth]{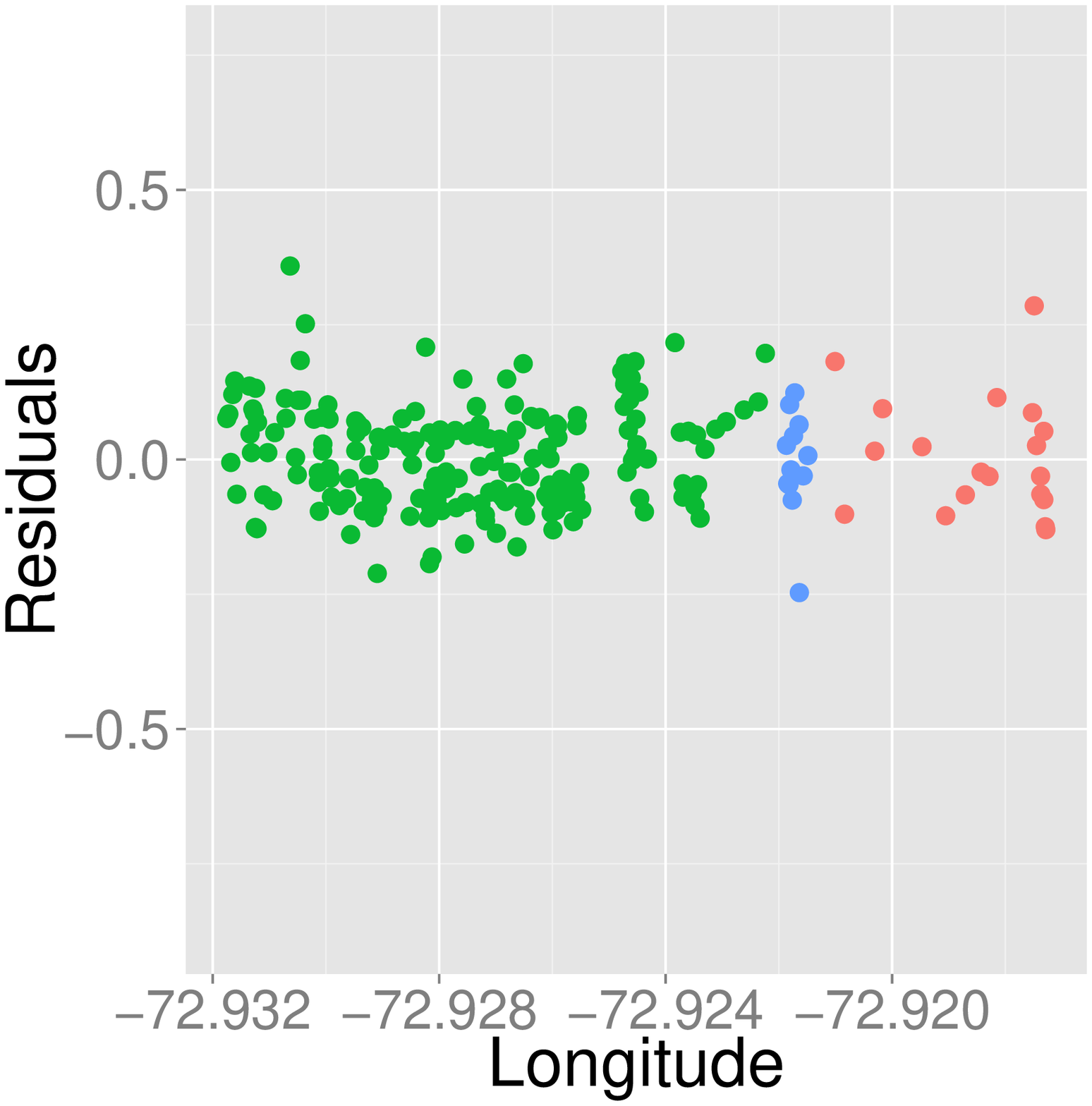}
    \caption{}
  \end{subfigure}
  \caption{Residuals by longitude for the New Haven housing example: (a) residuals from a linear model ($SE=0.200$) without any spatial or change point structure; (b) residuals from a linear model that estimates a different intercept for each neighborhood as identified by the posterior modal partition presented in Figure \ref{fig:newhavenhousing} ($SE=0.100$); (c) residuals from our method ($SE=0.092$). Residuals are colored for the three neighborhoods in the modal partition.}
\label{fig:newhavenhousing-resids}
\end{center}
\end{figure}

  For convenience, our subsequent discussions focus on the $\alpha=0.1$ results.
  In Figure \ref{fig:newhavenhousing-resids}, we compare the residuals from three different models: (a) a single linear model without any spatial component, (c) our method, which allows for different coefficients within different neighborhoods, and (b) an intermediate model sharing characteristics of the previous two, additionally using the modal partition from our method as a categorical variable in the linear model (a). All models use the same predicting variables with the same transformations, but (b) and (c) also indirectly use the adjacency structure which represents the spatial locations of the houses. The residuals from the original linear model (a) vary greatly by longitude, suggesting that some spatial effect has not been incorporated. The revised linear model (b) accounts for some of the spatial effect, and has a much smaller spread in the residuals. Our method produces the smallest standard deviation in the residuals among the three approaches, and the associated plot for (c) shows the residuals to be far more homogeneous across longitudes. 

  One might argue that a minimum spanning tree over all houses may not be the most intuitive graph structure for modeling the data. We concede there are alternative approaches. A $k$-nearest neighbors approach would require defining $k$ beforehand and even for small values of $k$, such as $k=4$, would result in joining some houses that are very far apart. Another possible approach is to create the graph structure using street names and numbers to join adjacent houses; however, the closest house may lie around the corner on a different street. We chose to use a minimum spanning tree because among these choices, it is the most objective way of generating the graph.
\section{Discussion} 
\label{sec:discussion}
This article proposes a Bayesian methodology that addresses a general class of change point problems. We demonstrated its use and performance on multivariate change point analysis for sequential data, univariate change point analysis for data on a grid graph, and linear regression change point analysis for data on a general graph. Our approach performs well on real data examples -- Quebec streamflow data and New Haven real estate data. For the Quebec streamflow problem, our Bayesian methodology confirmed the change point discovered by \cite{perreault2000retrospective} and provided additional information about other possible change points. For the New Haven real estate problem, our method identified a modal partition that was consistent with our \textit{a priori} knowledge of the neighborhoods in that area of New Haven.

We compared the performance of our method to BCP-Grid \citep{barryimage} using a diverse set of 20 simulated scenes on a grid graph. Our BCP-Graph algorithm with pseudo-APPs should be used with caution when many island nodes are suspected in the data. However, we showed that the BCP-Graph algorithms are more robust than BCP-Grid to the choice of the tuning parameter $\alpha$, especially when using pseudo-APPs; the original BCP-Grid algorithm exhibited extreme sensitivity to the choice of $\alpha$, as noted in \cite{barryimage}, and is therefore not recommended. In general, the use of pseudo-APPs helps improve performance by preventing the formation of many island nodes. As a result, the use of pseudo-APPs increases the useable range of $\alpha$ when working on real data problems where many island nodes are not expected. 

\newpage
\appendix

\makeatletter   
 \renewcommand{\@seccntformat}[1]{APPENDIX~{\csname the#1\endcsname}.\hspace*{1em}}
 \makeatother

\section{Selected Derivations}
\label{sec:appendix}
  The within-block density, shown below, is dependent upon $\tau_S$.
    If $\tau_S=0$, then \cite{barry93} gives
    \begin{align}
    &f(y_S|\bm{x_S}, \alpha_0, \sigma^2, \bm{w}, \tau_S=0)  
     \propto\left(\frac{1}{\sigma^2}\right)^{n_S/2} \exp\left\{-\frac{n_S w_0(\bar y_S- \alpha_0)^2+\sum_{i\in S}(y_{i}-\bar y_S)^2}{2 \sigma^2}\right\}.
    \end{align}
    If instead $\tau_S=1$, then we fit a linear model within block $S$:
    \begin{align}
    &f(y_S|\bm{x_S}, \bm {\gamma_S}, \sigma^2, \tau_S=1)\\
    &\propto \left(\frac{1}{\sigma^2}\right)^{n_S/2}\exp\left\{-\frac{1}{2 \sigma^2}\left[y_S^\top y_S - 2 \bm {\gamma_S}^\top\widetilde X_S^\top y_S + \bm {\gamma_S}^\top \widetilde X_S^\top \widetilde X_S \bm {\gamma_S}\right]\right\}\\
    &= \left(\frac{1}{\sigma^2}\right)^{n_S/2}\exp\left\{-\frac{1}{2 \sigma^2}\left[y_S^\top y_S - 2\bm {\hat \gamma_S}^\top \widetilde X_S^\top y_S  + \bm {\hat \gamma_S}^\top\widetilde X_S^\top y_S  + \bm {\hat \gamma_S}^\top \widetilde X_S^\top y_S - 2 \bm {\gamma_S}^\top\widetilde X_S^\top y_S + \bm {\gamma_S}^\top \widetilde X_S^\top \widetilde X_S \bm {\gamma_S}\right]\right\}\\
    &= \left(\frac{1}{\sigma^2}\right)^{n_S/2}\exp\left\{-\frac{1}{2 \sigma^2}\left[(y_S-\widetilde X_S \bm {\hat \gamma_S})^\top(y_S-\widetilde X_S\bm {\hat \gamma_S}) + (\bm {\hat \gamma_S} - \bm {\gamma_S})^\top \widetilde X_S^\top \widetilde X_S (\bm {\hat \gamma_S}- \bm {\gamma_S})\right]\right\},\;\;\;\;\label{eq:bcpr-inblock1}
    \end{align}
    where $\bm{\hat \gamma_S}$ is the ordinary least squares (OLS) estimate of the coefficient vector $\bm{\gamma_S}$. We write $\bm{\hat\gamma_S}=(\hat \alpha_S, \bm{\widetilde{ \beta}_S})$ where $\hat \alpha_S = \bar y_S$ and $\bm{\widetilde{\beta}_S}$ are the OLS estimates of the intercept and non-intercept components; we distinguish the OLS estimate $\bm{\widetilde\beta_S}=\left[(\widetilde X_S^\top \widetilde X_S)^{-1} \widetilde X_S^\top y_S\right]_{-1}$ from the Bayes estimate $\bm{\hat \beta_S}=\left[(\widetilde{X}_S^\top\widetilde{X}_S+Z_S^{-1})^{-1}\widetilde{X}_S^\top y_S\right]_{-1}$. We consider the last term in the exponent in \eqref{eq:bcpr-inblock1} and expand out $\bm {\hat \gamma_S}$ in terms of its components:
    \begin{align}
    &(\bm {\hat \gamma_S}- \bm{\gamma_S})^\top\widetilde X_S^\top \widetilde X_S (\bm {\hat \gamma_S}- \bm{\gamma_S})\\
    &=n_S(\alpha_S-\hat \alpha_S)^2  + (\bm{\beta_S} - \bm{\widetilde\beta_S})^\top {\widetilde X_S^\top \widetilde X_S}_{(-1,-1)} (\bm{\beta_S} - \bm{\widetilde\beta_S})\\
    &=n_S(\alpha_S-\bar y_S)^2+ (\bm{\beta_S} - \bm{\widetilde\beta_S})^\top {\widetilde X_S^\top \widetilde X_S}_{(-1,-1)} (\bm{\beta_S} - \bm{\widetilde\beta_S}).
    \end{align}
    Since $\alpha_S \sim N(\alpha_0, \sigma_0^2/n_S)$, we average \eqref{eq:bcpr-inblock1}  over $\alpha_S$ to get
    \begin{align}
    &f(y_S|\bm{x_S}, \alpha_0, w_0, \bm{\beta_S}, \sigma^2, \tau_S=1)\\
    &\propto \left(\frac{1}{\sigma^2}\right)^{n_S/2}\exp\left\{-\frac{1}{2 \sigma^2}\left[n_S w_0(\bar y_S- \alpha_0)^2 + \sum_{i\in S} (y_i-\bar y_S)^2 + (\bm{\beta_S} - \bm{\widetilde\beta_S})^\top {\widetilde X_S^\top \widetilde X_S}_{(-1,-1)} (\bm{\beta_S} - \bm{\widetilde\beta_S})\right]\right\}.
    \end{align}
    We rewrite the prior on $\bm \beta_S$ as $N_k(\bm 0, \sigma^2 Z_{S(-1,-1)})$ and average over $\bm \beta_S$ to arrive at
    \begin{align}
    &f(y_S|\bm{x_S}, \alpha_0, \sigma^2, \bm{w}, \tau_S=1)\\
     \propto&\left(\frac{1}{\sigma^2}\right)^{n_S/2} \left|{(\widetilde{X}_S^\top\widetilde{X}_SZ_S + I)}_{(-1,-1)}\right|^{-1/2}\\
     &\times \exp\left\{-\frac{n_S w_0(\bar y_S- \alpha_0)^2+\sum_{i\in S}\sum_j(y_{ij}-\bar y_S)^2 - \bm{\hat\beta_S}^\top(\widetilde{X}_S^\top\widetilde{X}_S+Z_S^{-1})_{(-1,-1)}\bm{\hat\beta_S}}{2 \sigma^2}\right\}.
    \end{align}   

    The joint density of all observations is then
    \begin{align}
    &f(y|\bm{x}, \alpha_0, \sigma^2, \bm{w}, \tau, \rho) \\
      \propto&\left(\frac{1}{\sigma^2}\right)^{n/2} \prod_{S:\tau_S=1}\left|{(\widetilde{X}_S^\top\widetilde{X}_SZ_S + I)}_{(-1,-1)}\right|^{-1/2}\\
  &\times\exp\left\{-\frac{W+w_0 \sum n_S(\bar y_S - \alpha_0)^2 - \sum_{S:\tau_S=1}\bm{\hat\beta_S}^\top(\widetilde{X}_S^\top\widetilde{X}_S+Z_S^{-1})_{(-1,-1)}\bm{\hat\beta_S}}{2 \sigma^2}\right\}.
    \end{align}
    Averaging over $\alpha_0$, $\sigma^2$, and finally $w_0$, we get for $b>1$,
    \begin{align}
    f(y|\bm{x}, \rho, \bm{w}, \tau) \propto&
      \frac{w_0'^{(b-1)/2}\prod_{S:\tau_S=1}\left|(\widetilde{X}_S^\top\widetilde{X}_SZ_S + I)_{(-1,-1)}\right|^{-1/2}}{B^{(b+1)/2} \widetilde{W}^{(n-b-2)/2}}\\
      &\times Beta\left(\frac{Bw_0'/\widetilde{W}}{1+Bw_0'/\widetilde{W}}; \frac{b+1}{2}, \frac{n-b-2}{2}\right).
    \end{align}
    Again, we consider the two cases: $B=0$ versus $B>0$. For $B>0$,
        \begin{align}
        f(\rho, \tau|y, \bm{x}, \bm{w}) 
          \propto&  f(\tau|\rho)f(\rho) f(y|\bm{x}, \rho, \bm{w}, \tau) \\
          \propto&\prod_{S}\left[\left(\frac{d}{n_S+d}\mathbbm{1}\{n_S\geq 2k\} + \mathbbm{1}\{n_S<2k\}\right)\mathbbm{1}\{\tau_S=0\}\right.\\
          &\quad\quad\left.+\left(\frac{n_S}{n_S+d}\mathbbm{1}\{n_S\geq 2k\}\right)\mathbbm{1}\{\tau_S=1\}\right]\\
          &\times\alpha^{l(\rho)}  \\          
          &\times
          \frac{w_0'^{(b-1)/2}\prod_{S:\tau_S=1}\left|(\widetilde{X}_S^\top\widetilde{X}_SZ_S + I)_{(-1,-1)}\right|^{-1/2}}{B^{(b+1)/2} \widetilde{W}^{(n-b-2)/2}}\\
          &\times Beta\left(\frac{Bw_0'/\widetilde{W}}{1+Bw_0'/\widetilde{W}}; \frac{b+1}{2}, \frac{n-b-2}{2}\right).
        \end{align}
      It is not possible to average over $\bm{w}$ to get a closed form expression for $f(\rho, \tau|y, \bm{x})$; we can get a closed form expression for $f(\bm{w}|y, \bm{x}, \rho, \tau)$, which for $B>0$ is 
      \begin{align}
      f(\bm{w}|y, \bm{x}, \rho, \tau) 
        \propto& f(y|\bm{x}, \rho,\bm{w}, \tau) f(\bm{w}) \\
       \propto& \frac{w_0'^{(b-1)/2}\prod_{S:\tau_S=1}\left|(\widetilde{X}_S^\top\widetilde{X}_SZ_S + I)_{(-1,-1)}\right|^{-1/2}}{B^{(b+1)/2} \widetilde{W}^{(n-b-2)/2}}\\
          &\times Beta\left(\frac{Bw_0'/\widetilde{W}}{1+Bw_0'/\widetilde{W}}; \frac{b+1}{2}, \frac{n-b-2}{2}\right).
    \end{align}
    If instead $B=0$, then
    \begin{align}
    f(\rho, \tau|y, \bm{x}, \bm{w}) 
          \propto&\prod_S\left[\left(\frac{d}{n_S+d}\mathbbm{1}\{n_S\geq 2k\} + \mathbbm{1}\{n_S<2k\}\right)\mathbbm{1}\{\tau_S=0\}\right.\\
          &\quad\quad\left.+\left(\frac{n_S}{n_S+d}\mathbbm{1}\{n_S\geq 2k\}\right)\mathbbm{1}\{\tau_S=1\}\right]\\
          &\times\alpha^{l(\rho)}  \\          
          &\times
          \frac{w_0'\left|(\widetilde{X}_S^\top\widetilde{X}_SZ_S + I)_{(-1,-1)}\right|^{-1/2}}{ \widetilde{W}^{(n-1)/2}}\;\;\;\;\;\mbox{ and}\\
      f(\bm{w}|y, \bm{x}, \rho, \tau) 
       \propto& \frac{w_0'\left|(\widetilde{X}_S^\top\widetilde{X}_SZ_S + I)_{(-1,-1)}\right|^{-1/2}}{\widetilde{W}^{(n-1)/2}}.
    \end{align}    

\section{Supplementary Materials}
\label{sec:supplementary}
\begin{figure}
  \begin{center}  
      \includegraphics*[width=.9\textwidth]{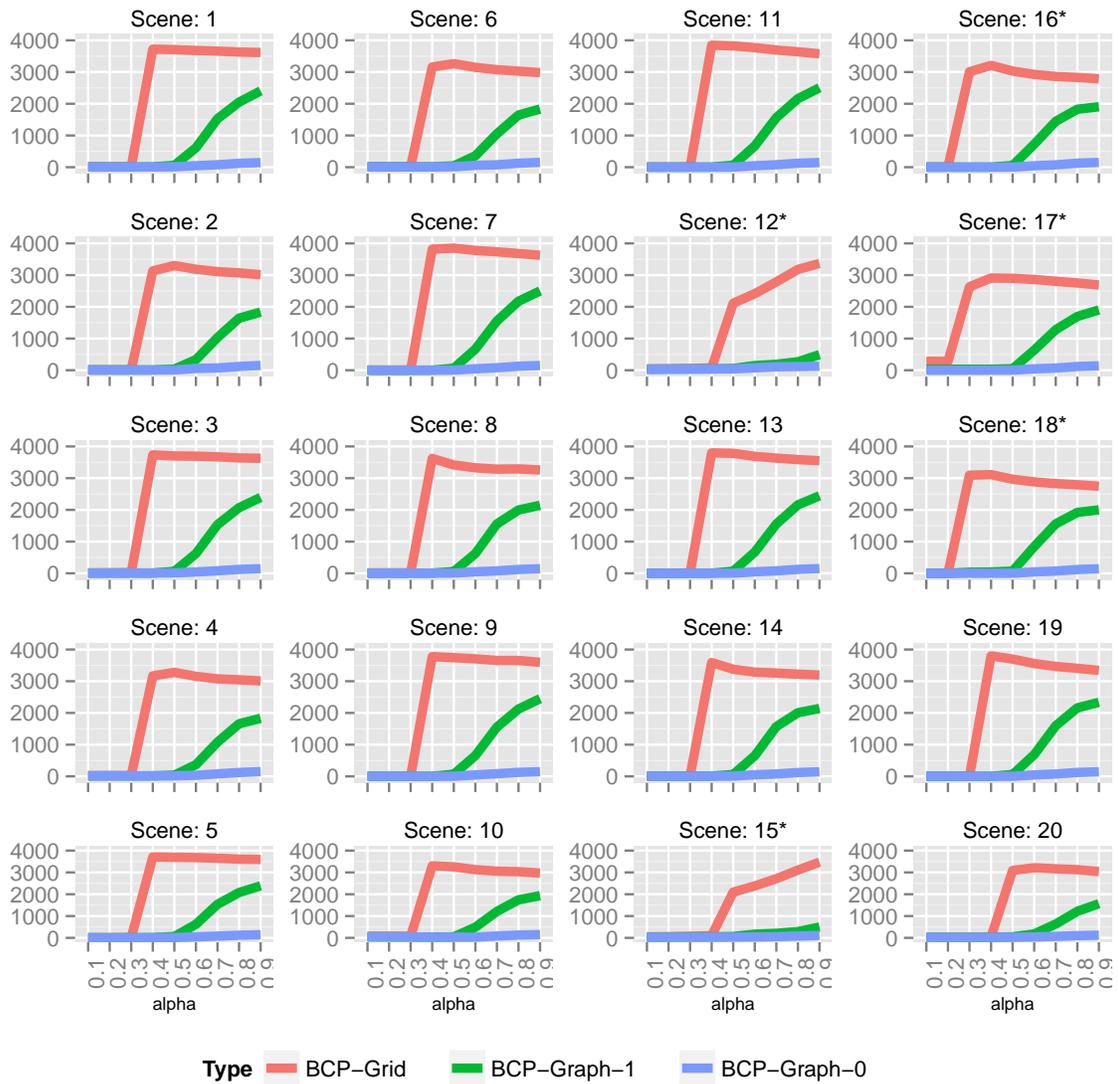}
    \caption{Mean runtime (seconds) for all simulated grid scenes. Scenes marked with asterisks generally attain lower MSEs at higher values of $\alpha$.}
  \label{fig:gridscenes-runtime}
  \end{center}
\end{figure} 
\begin{figure}
  \begin{center}  
      \includegraphics*[width=.9\textwidth]{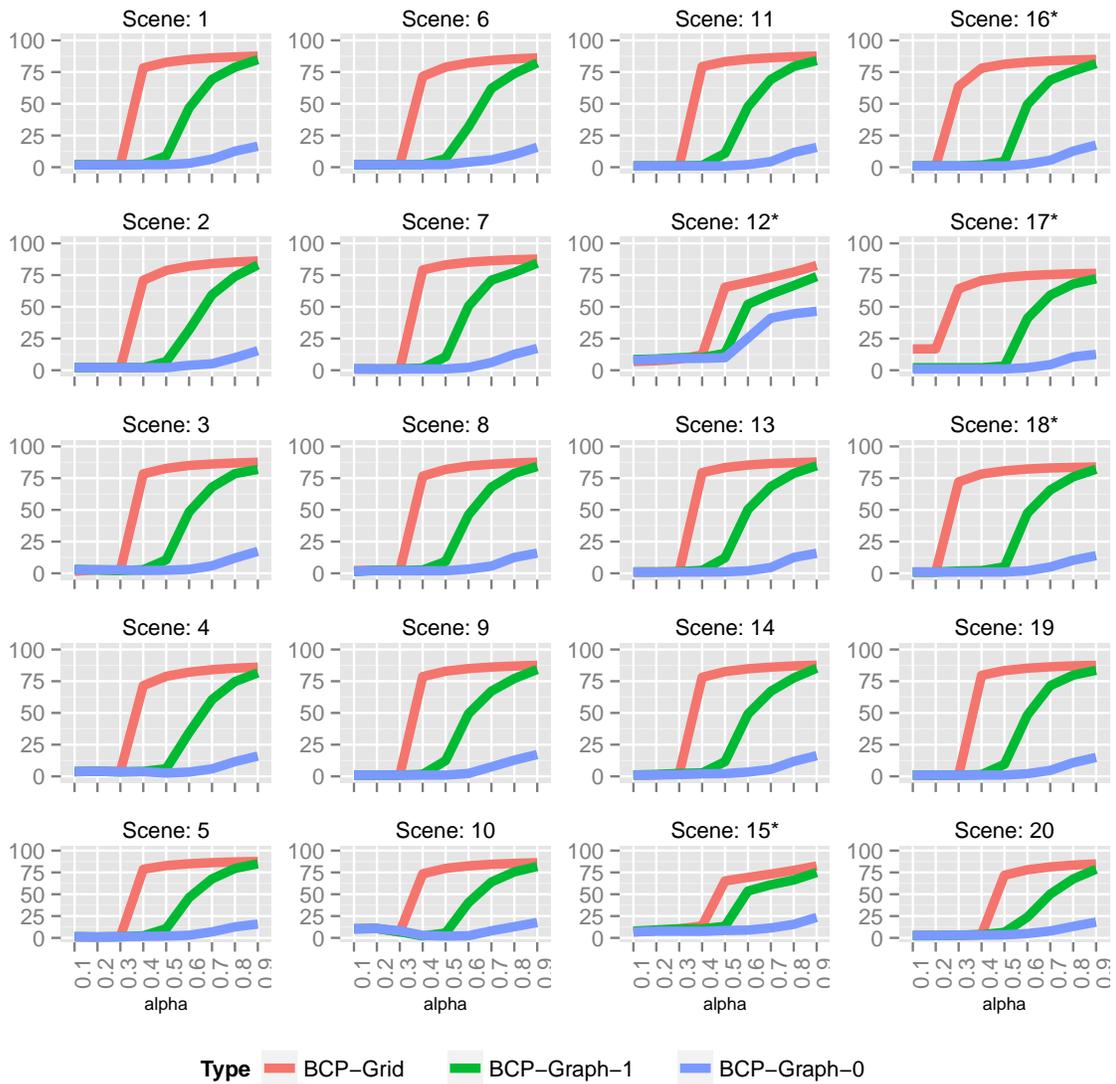}
    \caption{Mean number of blocks per iteration for all simulated grid scenes. Scenes marked with asterisks generally attain lower MSEs at higher values of $\alpha$.}
  \label{fig:gridscenes-meanblocks}
  \end{center}
\end{figure} 
\newpage
\bibliographystyle{asa} 
\bibliography{refs}

\end{document}